\documentclass[11pt]{article}
\usepackage{amsmath,color}
\usepackage{amssymb}
\usepackage{amsfonts}
\usepackage{graphicx}
\usepackage{color}

\definecolor{darkgreen}{rgb}{0,0.35,0}

\numberwithin{equation}{section}

\input{tcilatex}

\begin{document}

\title{\textbf{Minimal Log Gravity}}
\author{Gaston Giribet$^{1,2}$ \and Yerko V\'asquez$^3$}
\maketitle


\begin{center}
\textit{$^{1}$}Departamento de F\'{\i}sica, Universidad de Buenos Aires
FCEN-UBA and IFIBA-CONICET,

\textit{Ciudad Universitaria, Pabell\'{o}n I, 1428, Buenos Aires, Argentina.}

\medskip

\textit{$^{2}$}Instituto de F\'{\i}sica, Pontificia Universidad Cat\'olica
de Valpara\'{\i}so,\textit{\ }

\textit{Casilla 4950, Valpara\'{\i}so, Chile.} \\[0pt]

\medskip

\textit{$^{3}$}Departamento de F\'{\i}sica, Universidad de la Serena,\textit{%
\ }

\textit{Avenida Cisternas 1200, La Serena, Chile.} \\[0pt]

\medskip

\medskip

\medskip
\end{center}

Minimal Massive Gravity (MMG) is an extension of three-dimensional
Topologically Massive Gravity that, when formulated about Anti-de Sitter
space, accomplishes to solve the tension between bulk and boundary unitarity
that other models in three dimensions suffer from. We study this theory at
the chiral point, i.e. at the point of the parameter space where one of the
central charges of the dual conformal field theory vanishes. We investigate
the non-linear regime of the theory, meaning that we study exact solutions
to the MMG\ field equations that are not Einstein manifolds. We exhibit a
large class of solutions of this type, which behave asymptotically in
different manners. In particular, we find analytic solutions that represent
two-parameter deformations of extremal Ba\~{n}ados-Teitelboim-Zanelli (BTZ)
black holes. These geometries behave asymptotically as solutions of the
so-called Log Gravity, and, despite the weakened falling-off close to the
boundary, they have finite mass and finite angular momentum, which we
compute. We also find time-dependent deformations of BTZ that obey
Brown-Henneaux asymptotic boundary conditions. The existence of such
solutions shows that Birkhoff theorem does not hold in MMG at the chiral
point. Other peculiar features of the theory at the chiral point, such as
the degeneracy it exhibits in the decoupling limit, are discussed.

\section{Introduction}

Minimal Massive Gravity (MMG), introduced in Ref. \cite{MMG}, is an
extension of three-dimensional Topologically Massive Gravity (TMG) which,
when formulated about Anti-de Sitter space (AdS), accomplishes solving the
tension between bulk and boundary unitarity. As TMG, MMG\ about AdS$_{3}$
propagates a single local degree of freedom \cite{TMG}. However, in contrast
to what happens in other models of three-dimensional gravity such as TMG or
New Massive Gravity (NMG) \cite{NMG}, in MMG\ the AdS$_{3}$ graviton
excitation happens to have positive energy for the same values of coupling
constants for which the central charges of the dual conformal field theory
(CFT) turn out to be positive. Therefore, MMG seems to solves this
bulk-boundary unitarity puzzle in an ingenious way \cite{MMG,MMG2}.

Having a minimal model of three-dimensional gravity that allows for positive
graviton excitations and, at the same time, positive central charges in the
dual theory, a natural question arises as to what happens with the graviton
excitation at the so-called chiral point, i.e. at the point of the parameter
space where one of the two central charges of the dual CFT$_{2}$ vanishes.
This question is natural as the physics of the graviton excitation at the
chiral point was the main point in the discussion about the consistency of
the so-called Chiral Gravity; see \cite{CG, DCWW, GJ, GKP, MSS}.

Being parity-odd theories, both TMG\ and MMG\ in AdS$_{3}$ have an
asymptotically symmetry algebra generated by two copies of Virasoro algebra
with different central charges, $c_{\pm }$. The difference between these
central charges, $c_{+}-c_{-}$, controls the diffeomorphism anomaly in the
dual CFT$_{2}$. There is a point (or a curve) in the parameter space where
one of these central charges (say $c_{-}$) vanishes. It is commonly believed
that at that point, and provided suitable boundary conditions are imposed,
the boundary theory becomes a chiral CFT$_{2}$. At the chiral point, the
bulk theory also exhibits peculiar features, like the appearance of new
solutions with different boundary conditions. Depending on the asymptotic
boundary conditions considered, the local degrees of freedom of the bulk
theory can vary. For instance, in the case of TMG at the chiral point, there
exist two different models that coexist, each of them exhibiting
substantially different features: One such theory is Chiral Gravity,
originally proposed by Li, Song and Strominger in Ref. \cite{CG}, which is
defined by considering TMG\ on AdS$_{3}$ at the point $c_{-}=0$ and imposing
Brown-Henneaux boundary conditions \cite{BH}. The other theory is the
so-called Log Gravity \cite{MSS}, which is defined by considering the same
action with the same values of the coupling constants, but demanding a
weakened version of the asymptotic boundary conditions originally proposed
by Grumiller and Johansson in Ref. \cite{GJ}. While the former theory is
conjectured to be dual to a chiral CFT$_{2}$, there is evidence suggesting
that the latter is dual to a non-unitary (logarithmic) CFT$_{2}$.

Unlike Chiral Gravity, Log Gravity exhibits a propagating degree of freedom
(a bulk graviton). In Ref. \cite{GJ}, the energy of this graviton was
computed and shown to be negative (for values of the couplings for which
\thinspace $c_{+}$ was positive). Besides, in Ref. \cite{GKP} it was shown
that such graviton causes a linear instability in the theory, making
necessary to go to second order in perturbation theory to capture its actual
asymptotic behavior \cite{MSS}. An important piece of information to confirm
that a theory with a propagating graviton with weakened falling-off at the
chiral point $c_{-}=0$ actually existed was given in Ref. \cite{GAY}, where
it was proven that exact solutions obeying the boundary conditions of \cite%
{GJ} but not obeying those of \cite{BH} actually exist. This permitted
bringing the discussion between Chiral Gravity versus Log Gravity beyond the
linear and next-to-linear level. One of the results of this paper is\ to
extend this analysis to the case of MMG. We will show that MMG\ at the
chiral point ($c_{-}=0$) admits exact solutions that are asymptotically AdS$%
_{3}$ in the sense proposed in Ref. \cite{GJ} but fail to obey
Brown-Henneaux boundary conditions. Some of the solutions we exhibit
correspond to a two-parameter deformation of the extremal Ba\~{n}%
ados-Teitelboim-Zanelli (BTZ) black holes \cite{BTZ}, with a deformation
that may behave asymptotically either as the graviton of Log Gravity or as a
Brown-Henneaux solution. We compute the conserved charges of these MMG\
solutions and show that, despite the weakened asymptotics, the charges are
finite. We also consider the question as to whether exact solutions exist
which, while obeying the stronger Brown-Henneaux asymptotics, happen not to
be solutions of three-dimensional General Relativity. The (non-)existence of
such solutions at the chiral point was an important ingredient in the
discussion of \cite{MSS} about the contributions to the Chiral Gravity
partition function. Solutions of this sort for the case of TMG were
subsequently found in Ref. \cite{dBCD}, and here we show how they can be
generalized and extended to MMG. In particular, this leads us to show that
in this theory Birkhoff theorem does not hold, in the sense that there exist
circularly symmetric vacuum solutions that are time-dependent. We find such
time-dependent solutions explicitly.

The paper is organized as follows: In Section 2, we briefly introduce MMG
theory in AdS$_{3}$. In Section 3, we study the theory at the chiral point
at non-linear level. That is, we study analytic solutions of the theory that
are not the simple extensions of General Relativity solutions. In
particular, we present a two-parametric deformation of the extremal BTZ\
solution which has non-vanishing mass and angular momentum at the chiral
point. These solutions are explicit examples of solutions that behaves
asymptotically as Log Gravity solutions. We also find analytic
time-dependent deformations of the extremal BTZ\ black hole, showing that
Birkhoff theorem does not hold in this model. Nevertheless, we show that
these time-dependent excitations carry zero conserved charges. We also find
a much more general set of solutions, with different asymptotics. In Section
4, we present our conclusions.

\section{Minimal Log Gravity}

\subsection{Minimal Massive Gravity}

MMG is an extension of TMG. It can be conveniently described in the
so-called first order formalism, i.e. in terms of the vielbein one-form $%
e^{a}=e_{\mu }^{a}dx^{\mu }$ and the spin connection one-form $\omega
^{ab}=\omega _{\mu }^{ab}dx^{\mu }$ (where Latin indices refer to the
tangent space). First, let us recall that TMG can also be described in such
a way by first considering the Einstein-Hilbert action with cosmological
constant and then adding to it the exotic Chern-Simons term for $\omega ^{ab}
$ plus a term that couples a Lagrange multiplier $\lambda _{a}$ to the
torsion two-form $T^{a}=de^{a}+\omega _{b}^{a}\wedge e^{b}$; see for
instance \cite{Olivera}. Then, the MMG\ is defined by augmenting the TMG\
action written in the first order formalism by also including a term that is
quadratic in $\lambda ^{a}$. In this way, the variation of the action with
respect to $\omega ^{ab}$ produces an algebraic equation for $\lambda ^{a}$,
while the equation of motion associated to the variation with respect to $%
e^{a}$ involves its covariant derivative $d\lambda ^{a}+\omega
_{b}^{a}\wedge \lambda ^{b}$; see also \cite{otro1}. This yields a set of
third order field equations that defines MMG. In the second order formalism,
the equations of motion of MMG read

\begin{equation}
\sigma G_{\mu \nu }+\Lambda g_{\mu \nu }+\frac{1}{\mu }C_{\mu \nu }=-\frac{%
\gamma }{\mu ^{2}}J_{\mu \nu },  \label{G}
\end{equation}%
where $G_{\mu \nu }$ is the Einstein tensor, $G_{\mu \nu }\equiv R_{\mu \nu
}-\frac{1}{2}Rg_{\mu \nu }$, and $C_{\mu \nu }$ is the Cotton tensor%
\begin{equation}
C_{\mu \nu }=\frac{\epsilon _{\mu }^{\,\,\,\,\rho \sigma }}{\sqrt{-g}}\nabla
_{\rho }S_{\sigma \nu },  \label{C}
\end{equation}%
with $S_{\mu \nu }$ being the Schouten tensor, $S_{\mu \nu }\equiv R_{\mu
\nu }-\frac{1}{4}Rg_{\mu \nu }$, and where the tensor $J_{\mu \nu }$ is
defined also in terms of $S_{\mu \nu }$ as follows

\begin{equation}
J_{\mu \nu }=\frac{1}{2}\frac{\epsilon _{\mu }^{\,\,\,\,\rho \sigma }}{\sqrt{%
-g}}\frac{\epsilon _{\nu }^{\,\,\,\,\tau \eta }}{\sqrt{-g}}S_{\rho \tau
}S_{\sigma \eta }.  \label{J}
\end{equation}

In (\ref{G}), $\Lambda $ is the cosmological constant, $\mu $ is a coupling
constant with mass dimension one, and $\gamma $ is an arbitrary
dimensionless constant. $\sigma $ stands for a coefficient that can be set
to $\sigma =\pm 1$ by rescaling the other coupling constants.

One can verify that (\ref{J}) can be written as%
\begin{equation}
J_{\mu \nu }=R_{\mu }^{\,\,\,\,\rho }R_{\rho \nu }-\frac{3}{4}RR_{\mu \nu }-%
\frac{1}{2}g_{\mu \nu }\left( R^{\rho \sigma }R_{\rho \sigma }-\frac{5}{8}%
R^{2}\right) .  \label{J2}
\end{equation}

Since $J_{\mu \nu }$ is of second order in the metric, the Lanczos-Lovelock
theorem implies that it can not be covariantly conserved; that is, the
quantity $\nabla _{\mu }J_{\nu }^{\mu }$ can not vanish identically. In
fact, one can show that 
\begin{equation}
\nabla _{\mu }J_{\nu }^{\mu }=\frac{\epsilon _{\nu }^{\,\,\,\,\rho \sigma }}{%
\sqrt{-g}}S_{\rho }^{\eta }C_{\eta \sigma },  \label{mmm}
\end{equation}%
and it does not vanish identically. However, $\nabla _{\mu }J_{\nu }^{\mu
}=0 $ actually holds on-shell, once equations of motion (\ref{G}) are
imposed. The trace of (\ref{G}) reduces to%
\begin{equation}
\sigma R-6\Lambda +\frac{\gamma }{\mu ^{2}}\left( R_{\mu \nu }R^{\mu \nu }-%
\frac{3}{8}R^{2}\right) =0,  \label{Tr}
\end{equation}%
which coincides with the trace of NMG field equations \cite{NMG} for a
graviton mass $m=$ $\mu /\sqrt{\gamma }$.

The fact that covariant derivative (\ref{mmm}) does not vanish identically
immediately suggests that the theory would present at least two problems.
First, a problem will emerge when trying to couple the theory to matter.
This issue was addressed in Ref. \cite{matter}, where the kind of matter
content to which MMG\ can actually be coupled consistently was investigated.
A second problem is related to the computation of conserved charges.
Conservation of the field equations is a crucial step in the derivation of
the conserved charge formulae. This issue was recently studied in Ref.\ \cite%
{Tekin:2014jna}, where a method to compute conserved charges in MMG was
proposed. This consists of an extension of the Abbott-Deser-Tekin (ADT)
method, adapted to the peculiar case of MMG. In this paper, we do not need
to face any of these two problems because, on the one hand, we will consider
the theory (\ref{G}) in vacuum and, on the other hand, we will employ the
definition of conserved charges worked out in \cite{Tekin:2014jna} which, as
we will see, in our case also leads to consistent results.

\subsection{Minimal Massive Gravity in AdS$_3$}

We are interested in the theory about AdS$_{3}$ space, which is obviously a
solution of (\ref{G}). Assuming $R_{\mu \nu }=-(2/\ell ^{2})g_{\mu \nu }$ in
(\ref{Tr}), one finds%
\begin{equation}
\ell ^{2}=\frac{1}{2\Lambda }\left( -\sigma \pm \sqrt{1-\gamma \Lambda /\mu
^{2}}\right) ,  \label{27}
\end{equation}%
which gives an expression for the effective cosmological constant $\Lambda _{%
\text{eff}}\equiv -1/\ell ^{2}$ in terms of the coupling constants of the
theory (provided $\mu ^{2}\geq \gamma \Lambda $).

The theory about AdS$_{3}$ is conjectured to be dual to a two-dimensional
conformal field theory with central charges given by\footnote{%
Here, we will adopt the notation of Ref. \cite{Tekin:2014jna}.} 
\begin{equation}
c_{\pm }=\frac{3\ell }{2G}\left( \sigma +\frac{\gamma }{2\mu ^{2}\ell ^{2}}%
\pm \frac{1}{\mu \ell }\right) ,  \label{28}
\end{equation}%
with a normalization that leads to the Brown-Henneaux result $c_{\pm }=3\ell
/(2G)$ valid for General Relativity \cite{BH} in the limit $\mu \rightarrow
\infty $, and reproduces the expression for TMG\ when $\gamma =0$.

The central charge $c_{\pm }$ vanishes at the point%
\begin{equation}
\mu ^{2}\ell ^{2}\sigma \pm \mu \ell +\gamma /2=0
\end{equation}%
of the parameter space. In the case $\gamma =0$ this reduces to the TMG\
chiral point $\mu \ell =\pm 1$. As for the case of TMG, we will refer to the
point $c_{-}=0$ as the chiral point of the theory. The theory has also
another interesting point at $\mu ^{2}=\gamma \Lambda $, but we will not
study it here.

\section{Non-linear solutions}

\subsection{\textit{pp}-wave solutions in AdS$_{3}$}

The first class of exact solutions we will consider is $pp$-wave solutions
in AdS$_{3}$ space, also known as AdS-waves. For TMG these solutions were
studied in Ref. \cite{Eloy}. In MMG, AdS-waves solutions were also studied
recently \cite{otro3}, and here we review them because they are useful
solutions to gain intuition about the theory.

Let us start with the AdS$_{3}$ metric written in Poincar\'{e} coordinates;
namely

\begin{equation}
ds_{0}^{2}=\frac{\ell ^{2}}{z^{2}}\left( -2dx^{+}dx^{-}+dz^{2}\right) ,
\label{AdS}
\end{equation}%
where $z\in \mathbb{R}_{\geq 0}$. In these coordinates, the boundary of the
space is located at $z=0$. These coordinates, with $x^{\pm }\in \mathbb{R}$,
describe the Poincar\'{e} patch of AdS$_{3}$ space.

For MMG to admit the AdS vacuum (\ref{AdS}), the AdS$_{3}$ radius $\ell $,
the cosmological constant $\Lambda $, and the couplings parameters $\mu $
and $\gamma $ must obey (\ref{27}) for $\ell ^{2}>0$. The scale of the
cosmological constant, $\ell _{\Lambda }\equiv |\Lambda |^{-1/2}$ and the AdS%
$_{3}$ radius $\ell $ coincide when $\mu \rightarrow \infty $ or $\gamma =0$.

The metrics of AdS$_{3}$-waves can be written as

\begin{equation}
ds^{2}=\frac{\ell ^{2}}{z^{2}}\left( -F(x^{+},z)\left( dx^{+}\right)
^{2}-2dx^{+}dx^{-}+dz^{2}\right) .  \label{ppwave}
\end{equation}%
These metrics are conformally-related to that of the \textit{pp}-waves.

The AdS$_{3}$-wave solutions (\ref{ppwave}) describe exact gravitational
waves propagating on an AdS$_{3}$ background; therefore, we have to consider
the value (\ref{27}) for the effective cosmological constant. With this
choice, the equation of motion reads%
\begin{equation}
\frac{1}{4\ell ^{4}\mu ^{3}z}\left( -(\gamma \mu +2\ell ^{2}\mu ^{3}\sigma
)\ell ^{2}\frac{\partial F}{\partial z}+(\gamma \mu +2\ell ^{2}\mu
^{3}\sigma )\ell ^{2}z\frac{\partial ^{2}F}{\partial z^{2}}-2\ell ^{3}\mu
^{2}z^{2}\frac{\partial ^{3}F}{\partial z^{3}}\right) =0.
\end{equation}

Considering a solution of the form $F\propto z^{\alpha }$, the
characteristic polynomial reads

\begin{equation}
\alpha \left( \alpha -2\right) \left( \ell ^{2}\gamma \mu -2\ell ^{3}(\alpha
-1)\mu ^{2}+2\ell ^{4}\mu ^{3}\sigma \right) =0,  \label{sssw}
\end{equation}%
and generic solution for the wave profile is

\begin{equation}
F(x^{+},z)=F_{+}(x^{+})\left( z/\ell \right) ^{1+\ell \mu \sigma +\frac{%
\gamma }{2\ell \mu }},  \label{solution1}
\end{equation}%
where the function $F_{+}(x^{+})$ only depends on the coordinate $x^{+}$.
The solutions $\alpha =0$ and $\alpha =2$ can be eliminated by coordinate
transformations.

In addition to the power law behavior (\ref{solution1}), new logarithmic
modes appear for some values of the parameters: For the special point of the
parameter space $\mu ^{2}\ell ^{2}\sigma +\mu \ell +\gamma /2=0$, such a
logarithmic solution exists and is given by

\begin{equation}
F(x^{+},z)=F_{+}(x^{+})\log (z).
\end{equation}

Analogously, for the special point $\mu ^{2}\ell ^{2}\sigma -\mu \ell
+\gamma /2=0$, the logarithmic solution is given by

\begin{equation}
F(x^{+},z)=F_{+}(x^{+})\left( z^{2}\log (z)-z^{2}/2\right) .
\end{equation}

In these logarithmic solutions we have discarded the constant term and the
quadratic term in the variable $z$, because, as said, can be eliminated by
coordinate transformations.

The appearance of logarithmic AdS$_{3}$-waves solutions suggests that much
more general solutions with such near-boundary behavior exist in the theory
at the chiral point. In the next subsections we will see that this is indeed
the case. For instance, such as one can act on AdS$_{3}$ space with global
transformations and generate in this way interesting causal structures, like
black holes, there exists a rich variety of solutions that are locally
equivalent to AdS$_{3}$-waves discussed above and exhibit interesting
properties.

\subsection{Ansatz for non-linear deformation of AdS$_{3}$}

Let us begin by considering the vacuum metric (\ref{AdS}) and a deformation
of the form 
\begin{equation}
ds^{2}=ds_{0}^{2}+H_{-}(t,r)\left( dx^{-}\right) ^{2}+H_{+}(t,r)\left(
dx^{+}\right) ^{2},  \label{cero}
\end{equation}%
where $H_{\pm }(t,r)$ are two functions that may depend on time and the
radial direction (with coordinates $r\equiv \sqrt{2}\ell ^{2}/z$ and $x^{\pm
}\equiv t\pm \ell \phi $). Notice that here we are preserving the circular
symmetry but allowing for non-stationary solutions. In fact, we will see
that time-dependent solutions for $H_{\pm }(t,r)$ exist.

In terms of coordinates $r=\sqrt{2}\ell ^{2}/z$, $t=(x^{+}+x^{-})/2$ and $%
\phi =(x^{+}-x^{-})/(2\ell )$, metric (\ref{cero}) takes the form%
\begin{equation}
ds^{2}=-\frac{r^{2}}{\ell ^{2}}dt^{2}+\frac{\ell ^{2}}{r^{2}}%
dr^{2}+r^{2}d\phi ^{2}+H_{-}(t,r)(dt-\ell d\phi )^{2}+H_{+}(t,r)(dt+\ell
d\phi )^{2}.  \label{H}
\end{equation}

This turns out to be a particularly convenient ansatz to solve field
equations (\ref{G}). Inserting the form (\ref{H}) in the MMG\ equations of
motion, one obtains coupled third order differential equation for functions $%
H_{\pm }(t,r)$ that, despite the complexity of the higher-curvature terms,
in some cases can be solved analytically. Here, we are interested in the
chiral point $c_{-}=0$, namely in the case when $\mu ^{2}\ell ^{2}\sigma
-\mu \ell +\gamma /2=0$. At this point, a solution to the MMG\ equations of
motion with the form 
\begin{equation}
H_{-}(r)=2k\log \left( r\right) +k_{0}  \label{log}
\end{equation}%
arises;\ with $H_{+}$ being zero and with $k$ and $k_{0}$ being two
arbitrary constants. It is easy to check that this solution, which is the
simplest case of a much more general class we will studied below, behaves
asymptotically (i.e. at large $r$) as the Log Gravity excitations \cite{GJ,
MSS}, representing a non-linear realization of the theory. Besides, also at
the point $c_{-}=0$, one finds time-dependent solutions of the form 
\begin{equation}
H_{-}(r,t)=\tilde{k}\ t-\frac{\tilde{k}^{2}k_{\gamma }}{r^{4}}+k_{0},
\label{t}
\end{equation}%
with $H_{+}=0$, where $\tilde{k}$, $k_{0}$ and $k_{\gamma }$ are constants.
While $\tilde{k}$ and $k_{0}$ are arbitrary, constant $k_{\gamma }$ is
determined in terms of $\gamma $ and $\ell $ in a precise way we will
describe below. Deformation (\ref{t}) provides a time-dependent solution of
the MMG\ equations of motion. It is worthwhile noticing that, in contrast to
(\ref{log}), (\ref{t}) behaves asymptotically respecting Brown-Henneaux
boundary conditions. It can be shown that, despite being time-dependent,
this solution carries vanishing mass; see (\ref{329}) below.

Also at $c_{-}=0$, one finds solutions that deform the AdS$_{3}$ asymptotic
in a much more drastic way. For instance, one finds%
\begin{equation}
H_{+}(r)=2\hat{k}\ r^{2}\log \left( r\right) +k_{0},
\end{equation}%
with $H_{-}=0$ and with $\hat{k}$ being an arbitrary constant. Of course,
this latter solution has its mirror image $H_{-}(r)=2\hat{k}\ r^{2}\log
\left( r\right) +k_{0}$ with $H_{+}=0$ at $c_{+}=0$ (i.e. when $\ell ^{2}\mu
^{2}\sigma +\ell \mu +\gamma /2=0$.)

In the next subsections, we will see how these solutions can be generalized
and, in particular, lead to a two-parameter deformation of the extremal BTZ\
black hole.

\subsection{Logarithmic deformations of BTZ solution}

Let us begin by considering the extremal BTZ black hole solution;\ namely

\begin{equation}
ds_{\text{eBTZ}}^{2}=-N^{2}\left( r\right) dt^{2}+\frac{dr^{2}}{N^{2}\left(
r\right) }+r^{2}\left( N_{\phi }\left( r\right) dt-d\phi \right) ^{2},
\label{BTZ}
\end{equation}%
with the metric functions%
\begin{equation}
N^{2}\left( r\right) =\frac{r^{2}}{\ell ^{2}}-M+\frac{M^{2}\ell ^{2}}{4r^{2}}%
,\qquad N_{\phi }=\frac{M\ell }{2r^{2}},  \label{BTZ2}
\end{equation}%
and with $M$ being an integration constant. This solution, which for $M>0$
represents a maximally rotating black hole in three dimensions, solves
Einstein equations with negative $\Lambda $ in three dimensions \cite{BTZ}
and, then, it also solves the MMG\ field equations provided $\ell ^{2}>0$.
The black hole horizon is located at $r_{H}=\sqrt{M\ell ^{2}/2}$. At the
chiral point $\ell ^{2}\mu ^{2}\sigma -\ell \mu +\gamma /2=0$, solution (\ref%
{BTZ})-(\ref{BTZ2}) has vanishing mass and vanishing angular momentum \cite%
{Tekin:2014jna}.

Now, consider a deformation of the form%
\begin{equation}
ds^{2}=ds_{\text{eBTZ}}^{2}+N_{k}^{2}\left( r\right) \left( dt-\ell d\phi
\right) ^{2}.  \label{logmetric}
\end{equation}%
The case $M=0$ in (\ref{BTZ})-(\ref{logmetric}) corresponds to $%
H_{-}(r)=N_{k}^{2}\left( r\right) $, $H_{+}=0$ in (\ref{H}). Remarkably, at $%
\ell ^{2}\mu ^{2}\sigma -\ell \mu +\gamma /2=0$ the MMG\ field equations
admit the following configuration as an exact solution for arbitrary $M$,%
\begin{equation}
N_{k}^{2}\left( r\right) =k\log \left( \left( r^{2}-M\ell ^{2}/2\right)
/r_{0}^{2}\right) ,  \label{Nk}
\end{equation}%
where $k$ and $r_{0}$ are two arbitrary constants. Notice that (\ref{Nk})
reduces to (\ref{log}) in the case $M=0$. Asymptotically, (\ref{Nk}) also
behaves as a Log Gravity solution, as it damps-off slower than
Brown-Henneaux configurations \cite{BH} while obeys the Grumiller-Johansson
weakened boundary conditions \cite{GJ}. More precisely, solution (\ref{Nk})
behaves at large $r$ as follows%
\begin{eqnarray}
g_{tt} &\simeq &\frac{r^{2}}{\ell ^{2}}+\mathcal{O}(\log (r)),\qquad
g_{rr}\simeq \frac{\ell ^{2}}{r^{2}}+\mathcal{O}(r^{-4}), \\
g_{t\phi } &\simeq &\mathcal{O}(\log (r)),\qquad g_{\phi \phi }\simeq r^{2}+%
\mathcal{O}(\log (r)),
\end{eqnarray}%
where $\mathcal{O}(r^{-n})$ stands for functions of $t$ and $\phi $ whose
large $r$ behavior damps-off faster or equal than $1/r^{n}$. The presence of 
$\mathcal{O}(\log (r))$ is a typical feature that three- and
higher-dimensional higher-curvature gravity theories exhibit at critical
points of the moduli space.

Solution (\ref{BTZ})-(\ref{Nk}) represents a one-parameter deformation of
the extremal BTZ\ solution. As we will see in the next subsection, it
carries non-zero conserved charges, which are given in terms of the
integration constant $k$. The case $k=0$ reduces to the extremal BTZ\
solution. Even for $k\neq 0$, all the curvature scalars associated to this
solution are constants and independent of $k$; nevertheless, the solution
exhibits a pathology at $r=\sqrt{M\ell ^{2}/2}$, where the horizon of the
case $k=0$ is located. At that radius, the effective potential of geodesics
becomes infinite, tending either to $+\infty $ or $-\infty $ depending on
the sign of $k$. For $k\neq 0$, components $g_{tt}$, $g_{t\phi }$ and $%
g_{\phi \phi }$ of the metric blow up at the radius $r=\sqrt{M\ell ^{2}/2}$,
where the \textit{would be horizon} is located. The analysis of the geodesic
equations shows that angular velocity tends to infinity as the particles
approach that radius.

The solution (\ref{BTZ})-(\ref{Nk}), which has isometry group $SO(2)\times 
\mathbb{R}$, is not locally AdS$_{3}$ if $k\neq 0$. In fact, it is not even
conformally flat.

At the point $\ell ^{2}\mu ^{2}\sigma +\ell \mu +\gamma /2=0$ we obtain also
a deformation of BTZ\ with the form

\begin{equation}
\widehat{N}_{k}^{2}\left( r\right) =k\left( r^{2}-M\ell ^{2}/2\right) \log
\left( \left( r^{2}-M\ell ^{2}/2\right) /r_{0}^{2}\right) .  \label{318}
\end{equation}

This represents a much more drastic deformation of the AdS$_{3}$
asymptotics. However, close to the horizon it behaves much better that (\ref%
{Nk}), as (\ref{318}) vanishes in the limit $r\rightarrow \sqrt{M\ell ^{2}/2}
$.

\subsection{Conserved charges}

Now, let us compute the conserved charges of the solutions described above.
We will employ the method proposed by Tekin in Ref.\ \cite{Tekin:2014jna},
which provides a definition of conserved charges in MMG. To avoid
repetition, we will not review the details of the method of \cite%
{Tekin:2014jna} here; instead, we will refer to the original paper and to
the seminal works \cite{ADT1, ADT2}. Nevertheless, to facilitate the
discussion, we will work with the same notation as in \cite{Tekin:2014jna}.

Let us denote by $g_{\mu \nu }$ the spacetime metric and by $\bar{g}_{\mu
\nu }$ the \textit{background metric}, respect to which the charges will be
computed. These metrics have the same asymptotic Killing symmetries,
generated by $\bar{\xi}$, and are related by $g_{\mu \nu }\equiv \bar{g}%
_{\mu \nu }+h_{\mu \nu }$. Then, the charges are given by the following
formula,

\begin{equation}
Q^{\mu }\left( \bar{\xi}\right) =\frac{1}{2\pi G}\oint dl_{i}\left( \left(
\sigma +\frac{\gamma }{2\ell ^{2}\mu ^{2}}\right) q_{E}^{\mu i}(\bar{\xi})+%
\frac{1}{2\mu }q_{E}^{\mu i}(\bar{\Xi})+\frac{1}{2\mu }q_{C}^{\mu i}(\bar{\xi%
})\right) ,  \label{Q}
\end{equation}%
with the integral being evaluated on a circle at spatial infinity. Functions 
$q_{E}^{\mu i}(\bar{\xi})$, $q_{E}^{\mu i}(\bar{\Xi})$ and $q_{C}^{\mu i}(%
\bar{\xi})$ are defined in \cite{Tekin:2014jna} (see also Ref. \cite%
{Olmez:2005by}), and read

\begin{eqnarray}
q_{E}^{\mu i}(\bar{\xi}) &=&\sqrt{-\bar{g}}(\bar{\xi _{\nu }}\bar{\nabla}%
^{\mu }h^{i\nu }-\bar{\xi _{\nu }}\bar{\nabla}^{i}h^{\mu \nu }+\bar{\xi
^{\mu }}\bar{\nabla}^{i}h-\bar{\xi ^{i}}\bar{\nabla}^{\mu }h+h^{\mu \nu }%
\bar{\nabla}^{i}\bar{\xi}_{\nu }-  \notag \\
&&h^{i\nu }\bar{\nabla}^{\mu }\bar{\xi}_{\nu }+\bar{\xi}^{i}\bar{\nabla}%
_{\nu }h^{\mu \nu }-\bar{\xi}^{\mu }\bar{\nabla}_{\nu }h^{i\nu }+h\bar{\nabla%
}^{\mu }\bar{\xi}^{i}), \\
q_{C}^{\mu i}(\bar{\xi}) &=&\epsilon ^{\mu i\beta }\mathcal{G}_{\nu \beta }%
\bar{\xi}^{\nu }+\epsilon ^{\nu i\beta }\mathcal{G}_{\,\,\,\,\beta }^{\mu }%
\bar{\xi}_{\nu }+\epsilon ^{\mu \nu \beta }\mathcal{G}_{\,\,\,\,\beta }^{i}%
\bar{\xi}_{\nu },
\end{eqnarray}%
where $\bar{\Xi}^{\beta }=\epsilon ^{\alpha \nu \beta }\bar{\nabla}_{\alpha }%
\bar{\xi}_{\nu }/\sqrt{-\bar{g}}$ and where $\mathcal{G}_{\mu \nu }$ is the
linearized cosmological Einstein tensor

\begin{equation}
\mathcal{G}_{\mu \nu }=R_{\mu \nu }^{L}-\frac{1}{2}\bar{g}_{\mu \nu }R^{L}+%
\frac{2}{\ell ^{2}}h_{\mu \nu }
\end{equation}%
with the linearized Ricci tensor $R_{\mu \nu }^{L}$ given by

\begin{eqnarray}
R_{\mu \nu }^{L} &=&\frac{1}{2}(-\bar{\square}h_{\mu \nu }-\bar{\nabla}_{\mu
}\bar{\nabla}_{\nu }h+\bar{\nabla}^{\sigma }\bar{\nabla}_{\nu }h_{\sigma \mu
}+\bar{\nabla}^{\sigma }\bar{\nabla}_{\mu }h_{\sigma \nu }), \\
R^{L} &=&R_{\mu \nu }^{L}\bar{g}^{\mu \nu }+\frac{2}{\ell ^{2}}h=-\bar{%
\square}h+\bar{\nabla}_{\mu }\bar{\nabla}_{\nu }\bar{h}~^{\mu \nu }~+\frac{2%
}{\ell ^{2}}h,
\end{eqnarray}%
with $h=\bar{g}^{\mu \nu }h_{\mu \nu }$. Notice that all contractions and
raising and lowering indices must be done with the background metric $\bar{g}%
_{\mu \nu }$.

For a timelike Killing vector $\bar{\xi}^{\mu }=(-1,0,0)$, (\ref{Q})
corresponds to the energy, while for the spacelike Killing vector $\bar{\xi}%
^{\mu }=(0,0,1)$, it corresponds to the angular momentum. Then, choosing the
BTZ black hole with $M=0$ and $J=0$ as the background metric $\bar{g}_{\mu
\nu }$, we obtain the mass and the angular momentum corresponding to the
Killing vectors $\bar{\xi ^{\mu }}=-\partial _{t}$ and $\bar{\xi ^{\mu }}%
=\partial _{\phi }$, respectively. For the solution given by equations (\ref%
{logmetric})-(\ref{Nk}) at the chiral point $\ell ^{2}\mu ^{2}\sigma -\ell
\mu +\gamma /2=0$ , the conserved charges are given by

\begin{equation}
\mathcal{M}=\frac{2k}{\mu \ell G},\,\,\,\,\,\,\,\,\,\,\,\,\,\,\,\,\mathcal{J}%
=\frac{2k}{\mu G}.  \label{charges}
\end{equation}

As expected, in the limit $\gamma =0$ (\ref{charges}) agrees with the result
for TMG if the appropriate normalization of the action is considered; cf.
Eq. (3.51) of Ref. \cite{Olivera}; see also \cite{GAY}.

Then, what we have found here is an exact one-parameter deformation of the
extremal BTZ black hole that behaves asymptotically as a Log Gravity
solution and, besides, carries non-vanishing mass and angular momentum. In
the following subsections we will generalize this solution further.

\subsection{Time-dependent deformations}

Now, let us consider other exact solutions, which represent different type
of deformations of BTZ. Consider again the ansatz%
\begin{equation}
ds^{2}=-N^{2}\left( r\right) dt^{2}+\frac{dr^{2}}{N^{2}\left( r\right) }%
+r^{2}\left( N_{\phi }\left( r\right) dt-d\phi \right) ^{2}+\tilde{N}_{%
\tilde{k}}^{2}\left( t,r\right) \left( dt-\ell d\phi \right) ^{2},
\label{timesolution}
\end{equation}%
where now the deformation function depends on time. One can verify that at
the chiral point $\ell ^{2}\mu ^{2}\sigma -\ell \mu +\gamma /2=0$ the field
equations are solved for%
\begin{equation}
\tilde{N}_{\tilde{k}}^{2}\left( t,r\right) =\tilde{k}\ t-\frac{\tilde{k}%
^{2}\ell ^{6}}{N\left( r^{2}-M\ell ^{2}/2\right) ^{2}},  \label{Nqq}
\end{equation}%
where%
\begin{equation}
N=\frac{96}{5-4\ell \mu \sigma }=\frac{96}{1+2\gamma /(\ell \mu )}.
\label{rt}
\end{equation}

The case $M=0$ of (\ref{Nqq}) corresponds to (\ref{t}) with $\tilde{k}%
=k/\ell ^{2}$, $k_{\gamma }=\ell ^{6}/N$ . This time-dependent solution
generalizes one of the non-linear solutions of TMG\ found in Ref. \cite{dBCD}%
. Notice that, in fact, in the case $\ell \mu \sigma =1$ (i.e. $\gamma =0$)\
the particular case $M=0$ of (\ref{t}) reduces to Eq. (3.22) of Ref. \cite%
{dBCD}, for which $N=96$. It also generalizes solutions of TMG\ coupled to
NMG studied in Ref. \cite{Goya} to the case $M\geq 0$.

Solution (\ref{timesolution})-(\ref{rt}) presents a large $r$ behavior
consistent with the expansion%
\begin{eqnarray}
g_{tt} &\simeq &\frac{r^{2}}{\ell ^{2}}+\mathcal{O}(r^{0}),\qquad
g_{rr}\simeq \frac{\ell ^{2}}{r^{2}}+\mathcal{O}(r^{-4}), \\
g_{t\phi } &\simeq &\mathcal{O}(r^{0}),\qquad g_{\phi \phi }\simeq r^{2}+%
\mathcal{O}(r^{0}),
\end{eqnarray}%
so that it satisfies the Brown-Henneaux boundary conditions \cite{BH}.

The existence of this type of solutions in the case of TMG\ was relevant for
the discussion about the contribution to the partition function of Chiral
Gravity \cite{MSS}. This is because one of the assumptions in \cite{MSS} was
the non-existence of solutions obeying Brown-Henneaux boundary conditions
and not being Einstein manifolds. The discovery of such a solution in \cite%
{dBCD} proved that there exist non-Einstein contributions. Here, we have
shown that also in MMG\ in AdS$_{3}$ at the chiral point, solutions exist
that obey Brown-Henneaux asymptotic boundary conditions not being Einstein
manifolds.

On the other hand, the existence of time-dependent circularly symmetric
solution (\ref{timesolution})-(\ref{rt}) manifestly shows that in this
theory Birkhoff theorem does not hold. A natural question is what are the
conserved charges associated to this time-dependent deformation of extremal
BTZ. Notice that, despite being time-dependent, the linear dependence on
time appears in a next-to-leading term in the large $r$ expansion. One can
use the method of \cite{Tekin:2014jna} described in the previous subsection
and verify that the mass and angular momentum associated to (\ref%
{timesolution})-(\ref{rt})\ actually vanish; namely 
\begin{equation}
\mathcal{M}=0,\,\,\,\,\,\,\,\,\,\,\,\,\,\,\,\,\mathcal{J}=0.  \label{329}
\end{equation}

This raises the question as to whether non-Einstein solutions obeying
Brown-Henneaux conditions and having non-vanishing conserved charges
actually exist.

There is also another interesting point, which is where $N$ in (\ref{rt})
diverges;\ that is, when $\ell \mu \sigma =-\gamma \sigma =5/4$. There, a
solution appears which corresponds to replacing $\tilde{N}_{\tilde{k}%
}^{2}\left( t,r\right) $ in (\ref{Nqq}) by $\tilde{N}_{\tilde{k}}^{2}\left(
t,r\right) =k_{2}r^{2}+\tilde{k}t+k_{0}+N_{k}^{2}\left( r\right) $, with $%
N_{k}^{2}\left( r\right) $ given by (\ref{Nk}). Let us discuss this type of
solutions with more free parameters in the next subsection.

\subsection{Two-parameter deformations and generalizations}

Remarkably, if we \textit{add} solutions (\ref{logmetric}) and (\ref%
{timesolution}), a new solution to the field equations at the chiral point $%
\ell ^{2}\mu ^{2}\sigma -\ell \mu +\gamma /2=0$ is obtained;\ namely%
\begin{equation}
H_{-}(t,r)=N_{k}^{2}\left( r\right) +\tilde{N}_{\tilde{k}}^{2}\left(
t,r\right) ,
\end{equation}%
with $N_{k}^{2}(r)$ and $\tilde{N}_{\tilde{k}}^{2}\left( t,r\right) $
defined in equations (\ref{Nk}) and (\ref{Nqq}) also solves the MMG\ field
equations for arbitrary $k$ and $\tilde{k}$. We obtain in this way a
two-parameter deformation of extremal BTZ. The conserved charges of this
general solution can also be computed and shown to yield (\ref{charges}).

Besides, one can explore a much more general ansatz for $H_{\pm }$,
including dependence on the angular coordinate $\phi $, and still find
analytic solutions. For instance, perturbing the $M=0$ solution at the
chiral point $\mu ^{2}\ell ^{2}\sigma -\mu \ell +\gamma /2=0$, one finds that

\begin{equation}
ds^{2}=-\frac{r^{2}}{\ell ^{2}}dt^{2}+\frac{\ell ^{2}}{r^{2}}dr^{2}+\frac{%
r^{2}}{\ell ^{2}}dx^{2}+H_{-}(t,r,x)\ (dt-dx)^{2},
\end{equation}%
with $H_{+}=0$ and $H_{-}(t,r,x=\ell \phi )$ solves the field equations for
the following expansion%
\begin{equation}
H_{-}(t,r,x)=h_{-}^{(2)}(t,x)\ r^{2}+h_{-}^{(0)}(t,x)+h_{-}^{(0+)}\ \log
(r)+h_{-}^{(-4)}\ r^{-4}.
\end{equation}%
where $h_{-}^{(n)}(t,x)$ stand for functions of $t$ and $x$ that organize
the different terms of order $\mathcal{O}(r^{n})$, with $h_{-}^{(0+)}$ being
the coefficient of the logarithmic term. The explicit forms of these
functions are%
\begin{equation}
h_{-}^{(2)}(t,x)=k_{0}+k_{-}\ (t-x),\quad h_{-}^{(0)}(t,x)=\tilde{k}%
_{0}+k_{t}\ t+k_{x}\ x,
\end{equation}%
with constant coefficients%
\begin{equation}
h_{-}^{(0+)}=2k,\quad \quad h_{-}^{(-4)}=-\ell ^{6}(1+2\gamma /(\ell \mu
))(k_{x}+k_{t})^{2}/96.
\end{equation}%
where $k_{0}$, $k_{-}$, $\tilde{k}_{0}$, $k_{t}$, $k_{x}$, and $k$ are all
arbitrary constants. However, periodicity in the angular coordinate $\phi
\equiv x/\ell $ demands $k_{-}=k_{x}=0$.

Solutions with different chirality (i.e. with $H_{+}\neq 0$ at $c_{-}=0$)
can also be found. For instance, one finds%
\begin{equation}
H_{+}(t,r,x)=h_{+}^{(2+)}\ r^{2}\log (r)+h_{+}^{(2)}(t,x)\
r^{2}+h_{+}^{(0)}(t,x)
\end{equation}%
and $H_{-}=0$, with $h_{+}^{(2+)}$ being an arbitrary constant, and 
\begin{equation}
h_{+}^{(2)}(t,x)=k_{t}\ t+k_{x}\ x+k_{0},\qquad h_{+}^{(0)}(t,x)=\tilde{k}%
_{0}+k_{+}\ (t+x),
\end{equation}%
where, again, $k_{t}$, $k_{x}$, $k_{0}$, $\tilde{k}_{0}$, and $k_{+}$ are
all arbitrary constants.

\subsection{Degeneracy and decoupling limit}

An interesting phenomenon occurs in the limit where the Cotton tensor
disapears from the equations of motion (\ref{G}). This corresponds to the
limit $\mu \rightarrow \infty $ and $\gamma /\mu \rightarrow \infty $,
keeping the ratio $m^{2}\equiv \mu ^{2}/\gamma $ fixed. In this limit,
central charges (\ref{28}) read 
\begin{equation}
c_{\pm }=\frac{3\ell }{2G}\left( \sigma +\frac{1}{2m^{2}\ell ^{2}}\right) 
\end{equation}%
and, therefore, the chiral point corresponds to $m^{2}\ell ^{2}\sigma =-1/2$%
. At this point, the ansatz%
\begin{equation}
ds^{2}=ds_{\text{eBTZ}}^{2}+H_{-}\left( r\right) \left( dt-\ell d\phi
\right) ^{2},  \label{vidal}
\end{equation}%
with $H_{-}\left( r\right) $ depending only on $r$, solves the field
equations for arbitrary $H_{-}\left( r\right) $. This is consistent with the
fact that equation (\ref{sssw}), when divided by $\mu ^{3}$, is
automatically satisfied for $m^{2}\ell ^{2}\sigma =-1/2$ in the $\mu
\rightarrow \infty $ limit. At this point of the parameter space, other
solutions also exhibit special features. For instance, time-dependent
solution (\ref{t}) disapears as $N$ tends to zero in the limit $\mu \sim
\gamma /\mu \rightarrow \infty $.

The type of degeneracy that (\ref{vidal}) exhibits, is a common feature that
higher-curvature theories present at special points of the moduli spaces
where symmetry enhancement phenomena happen. This could also be a symptom
indicating that the theory is not well defined in such limit. It would be
interesting to understand this better.

\section{Conclusions}

In this paper, we have studied MMG theory, which is the extension of TMG\
proposed in Ref. \cite{MMG} that solves the tension between bulk and
boundary unitarity, a problem that theories like TMG\ and NMG have.
Specifically, we studied MMG\ about AdS$_{3}$ space at the chiral point,
i.e. at the point of the parameter space where one of the central charges of
the boundary theory vanishes. We considered both Brown-Henneaux \cite{BH}
and Grumiller-Johansson \cite{GJ} boundary conditions, which in the case of
TMG\ at the chiral point lead to the definition of Chiral Gravity \cite{CG}
and Log Gravity \cite{MSS} theories, respectively.

We studied the theory at non-linear level;\ that is, we studied exact
solutions to MMG\ field equation analytically, focusing our attention to
solutions of MMG equations that are not Einstein manifolds. Then, we found
two-parametric deformations of extremally rotating BTZ\ black holes. In
particular, we found exact solutions that behave asymptotically as Log
Gravity excitations. We computed the conserved charges of such
configurations and showed that, despite the weakened AdS$_{3}$ asymptotic,
these Log deformations of BTZ black holes exhibit finite mass and finite
angular momentum, which we compute.

We also found time-dependent deformations of extremal BTZ that depend on
time linearly. These solutions do respect the stronger AdS$_{3}$ boundary
conditions, and it implies that at the chiral point solutions exist that
behave asymptotically as Brown-Henneaux gravitons despite not being Einstein
manifolds. Despite being time-dependent, these solutions have vanishing
conserved charges. This opens the question about the contribution of such
type of configurations to the partition function of the theory.

The general case we considered includes both Log Gravity and time-dependent
deformations, representing two-parameter deformations of BTZ. This led us to
show explicitly that Birkhoff theorem does not hold in this theory. We also
investigate more general solutions that exist at the chiral point of MMG.
Besides, the theory formulated at other points of the parameter space also
exhibits interesting solutions of different types. For instance, it admits
locally AdS$_{2}\times \mathbb{R}$ geometries that are worth exploring.

\begin{equation*}
\end{equation*}%
This work was funded by CONICET\ Grant PIP\ 0595/13 and UBACyT
20020120100154BA (G.G.), and by Comisi\'{o}n Nacional de Ciencias y Tecnolog%
\'{\i}a through FONDECYT Grant 11121148 (Y.V.). G.G. thanks Pontificia
Universidad Cat\'{o}lica de Valpara\'{\i}so for the hospitality.

\end{document}